\documentstyle[graphics]{mn}
\newcommand{\kms}{{km s$^{-1}$}}
\newcommand{\smk}{{km$^{-1}$ s}}

\newcommand{\centm}{{cm$^{-2}$}}
\newcommand{\etal}{{et al.}~}

\newcommand{\bfk}{{\bf k}}

\newcommand{\bc}{\begin{center}}
\newcommand{\be}{\begin{equation}}
\newcommand{\ee}{\end{equation}}
\newcommand{\ec}{\end{center}}

\newcommand{\hydra}{{\sc{HYDRA }}}

\newcommand{\cmbfast}{{\sc {CMBFAST }}}
\newcommand{\vpfit}{{\sc {VPFIT}}}
\newcommand{\autovp}{{\sc {AUTOVP}}}
\newcommand{\lya}{Ly$\alpha$~}

\newcommand{\ie}{{i.e.~}}

\newcommand{\h} {{\rm H{\sc i}}}

\newcommand{\NHI}{\mbox{$N_{\rm H{\sc i}}$}}
\newcommand{\ltsima}{\mbox{$\; \buildrel < \over \sim \;$}}
\def \simlt{\lower.5ex\hbox{\ltsima}}            
\def \gtsima{\mbox{$\; \buildrel > \over \sim \;$}}
\def \simgt{\lower.5ex\hbox{\gtsima}}            

\title[Broadening of QSO \lya forest absorbers] 
{Broadening of QSO \lya forest absorbers}
\author[Theuns, Schaye \& Haehnelt]{Tom Theuns$^1$, Joop Schaye$^2$ 
and Martin G. Haehnelt$^1$\\
$^{\,1}$ Max-Planck Institut f\"ur Astrophysik, Postfach 123, 85740 Garching, Germany\\
$^{\,2}$ Institute of Astronomy, Madingley Road, Cambridge CB3 0HA, UK
}

\begin{document}
\maketitle

\begin{abstract}
We investigate the dependence of QSO \lya absorption features on the
temperature of the absorbing gas and on the amplitude of the underlying
dark matter fluctuations. We use high-resolution hydrodynamic
simulations in cold dark matter dominated cosmological models. In
models with a hotter intergalactic medium (IGM), the increased
temperature enhances the pressure gradients between low and high
density regions and this changes the spatial distribution and the
velocity field of the gas. Combined with more thermal broadening, this
leads to significantly wider absorption features in hotter models.
Cosmological models with little small scale power also have broader
absorption features, because fluctuations on the scale of the Jeans
length are still in the linear regime . Consequently, both the
amplitude of dark matter fluctuations on small scales and thermal
smoothing affect the flux decrement distribution in a similar way.
However, the $b$-parameter distribution of Voigt profile fits, obtained
by deblending the absorption features into a sum of thermally broadened
lines, is largely independent of the amount of small scale power, but
does depend strongly on the IGM temperature. The same is true for the
two-point function of the flux and for the flux power spectrum on small
scales. These three flux statistics are thus sensitive probes of the
temperature of the IGM. We compare them computed for our models and
obtained from a HIRES spectrum of the quasar Q1422+231 and conclude
that the IGM temperature at $z\sim 3.25$ is fairly high, $T_0\ga
15000$K. The flux decrement distribution of the observed spectrum is
fitted well by that of a $\Lambda$CDM model with that temperature.
\end{abstract}


\section{Introduction}

Neutral hydrogen in the intergalactic medium (IGM) produces a \lq
forest\rq~ of \lya absorption lines bluewards of the \lya emission line
in quasar spectra (Bahcall \& Salpeter 1965, Gunn \& Peterson
1965). The general properties of these \lya absorption lines are
remarkably well reproduced in hydrodynamic simulations of cold dark
matter (CDM) dominated cosmologies (Cen \etal 1994, Zhang, Anninos \&
Norman 1995, Miralda-Escud\'e \etal 1996, Hernquist \etal 1996, Wadsley
\& Bond 1996, Zhang \etal 1997, Theuns \etal 1998). In these
simulations, intergalactic gas settles into the potential wells of CDM
dominated sheets and filaments, aptly termed the cosmic web (Bond,
Kofman \& Pogosyan 1996). These large-scale density fluctuations in the
intergalactic gas are responsible for the weak absorption lines of
column density $\NHI \le 10^{14.5}$\centm~.

In the simulations, the typical widths of the peaks in the neutral
hydrogen density, which occur when a line-of-sight crosses a sheet or
filament, depend on the amplitude of dark matter fluctuations on small
scales and by the smoothing of these small scale fluctuations by gas
pressure (\lq Jeans smoothing\rq, e.g. Bryan \etal 1999). The widths of
the absorption troughs are then determined by differential Hubble flow
accross the spatially extended absorber (Hernquist \etal 1996),
peculair velocities (Hui 1999), and thermal broadening.

Widths of absorption features have traditionally been characterized by
the $b$-parameter distribution of Voigt profile fits, obtained by
deblending the absorption features into a sum of discrete absorption
lines. Recent high-resolution simulations have shown that these
absorption lines in a standard CDM cosmology have $b$-parameters
substantially narrower than observed in high-resolution Keck spectra
(Theuns \etal 1998, Bryan \etal 1999). Since the $b$-parameters have
been demonstrated to be at least partially thermally broadened
(Haehnelt \& Steinmetz 1998, Theuns \etal 1998, Hui \& Rutledge 1999,
Schaye et al. 1999a), Theuns \etal suggested that an increased
temperature of the IGM might resolve the discrepancy.

At the relevant redshifts ($2\la z\la 5$), the temperature of the low
density IGM is determined by the interplay between adiabatic cooling
and photoheating, which introduces a tight relation between density and
temperature which can be well approximated by a power law (Hui \&
Gnedin 1997).  Schaye et al. (1999a) used numerical simulations to
demonstrate that the observed cut-off in the Doppler parameter
distribution and its dependence on column density is the result of this
$\rho-T$ relation. The exact form of the relation depends on the
history of reionization and on cosmological parameters
(Miralda-Escud\'e \& Rees 1994, Hui \& Gnedin 1997). For a given
reionization history, higher temperatures are obtained for models with
a higher baryon density $\Omega_b h^2$ and a smaller expansion rate of
the Universe. Theuns \etal (1999) confirmed that in a model with high
baryon density $\Omega_b h^2=0.025$\footnote{We write the present day
Hubble constant as $H_0=100h$ \kms Mpc$^{-1}$}, low matter density
$\Omega_{\rm m}=0.3$ and $h=0.65$ the $b$-distribution of the simulated
absorption lines fits the observed distribution significantly better
than for the colder standard CDM model.

However, the value of $\Omega_b h^2$ required is higher than the
current best estimates $\Omega_b h^2=0.019$ (Burles \& Tytler 1998) and
it was suggested that additional heat sources neglected so far in the
numerical simulations might contribute significantly to the heating
rate. These include the photo electric heating by dust grains (Nath,
Sethi \& Shchekinov 1999) and Compton heating by the hard X-ray
background (Madau \& Efstathiou 1999). Alternatively, Abel \& Haehnelt
(1999) pointed out that the optically thin limit for the HeII
photoheating rate usually assumed in these simulations might lead to an
underestimate compared to the actual HeII photoheating rate by a factor
of $\sim 2-3$.

Here we use high resolution hydrodynamic simulations to investigate how
the IGM temperature and the amplitude of dark matter fluctuations
affect the distribution of the gas and influences the widths of
absorption features. The effect of different heating rates and the
influence of the amplitude of dark matter fluctuations on the gas
distribution are discussed in Section 3. In Section 4 we investigate
the relative importance of the three thermal smoothing effects, Jeans
smoothing, peculiar velocities and thermal broadening, and the
amplitude of the dark matter fluctuations on the $b$-parameter
distribution obtained by Voigt profile fitting. In Section 5 we discuss
the effects on the one- and two-point function of the flux and on the
flux power spectrum. Section 6 contains our conclusions.  As the
preparation of this paper was in its final stages a preprint by Bryan
\& Machacek (1999) was circulated that investigates some of the same
issues.

\section{Simulation parameters and preparation of mock spectra}

\label{sect:simulation}
\begin{table}
 \centering \begin{minipage}{70mm} \caption{Models simulated}
 \begin{tabular}{@{}llllllll@{}} 
Name	& $\Omega_m$	& $h$		& $\sigma_8$	&
 $\epsilon_{\rm He}$ & X-ray & $\log {T_0}$ & A \\[5pt]
L0.3	& 0.3     		& 0.65	& 0.9				& HM/3              & no   & 3.93 & 1.07 \\
L1	   & 0.3     		& 0.65	& 0.9				& HM                & no   & 4.07 & 0.92 \\
L2	   & 0.3     		& 0.65	& 0.9				& 2xHM              & no   & 4.19 & 0.78 \\
L3	   & 0.3     		& 0.65	& 0.9				& 3xHM              & no   & 4.26 & 0.68 \\
L1$\sigma .65$& 0.3  & 0.65	& 0.65			& HM	              & no   & 4.07 & 1.05 \\
L1$\sigma .4$& 0.3    & 0.65	& 0.4				& HM	              & no   & 4.07 & 1.25 \\
L3b	& 0.3     		& 0.65	& 0.9				& 3xHM              & no   & 4.26 & 0.59 \\
LX	   & 0.3     		& 0.65	& 0.9				& HM                & yes  & 4.14 & 0.81 \\
S3	   & 1.0     		& 0.50	& 0.5				& 3xHM              & no   & 4.25 & 0.78 \\

\label{table:runs}
\end{tabular}
\end{minipage}
\end{table}

We have simulated several different models, characterized by their
total matter density $\Omega_{\rm m}$, the value of the cosmological
constant $\Omega_\Lambda$, the rms of mass fluctuations in spheres of
8$h^{-1}$ Mpc, $\sigma_8$ and the heating rate. The parameters for
these models are summarized in Table~\ref{table:runs}. All these models
have $\Omega_m+\Omega_\Lambda=1$. For our reference model L1 we take
the currently favoured values $h=0.65$ (see Freedman \etal 1998),
$\Omega_b h^2=0.019$ (Burles and Tytler 1998), $(\Omega_{\rm
m},\,\Omega_\Lambda)=(0.3,0.7)$ (Efstathiou \etal 1999 and references
therein), $\sigma_8=0.9$ (Eke, Cole \& Frenk 1996) and a helium
abundance of $Y=0.24$ by mass. The IGM in this model is photoionized
and photoheated by the UV-background from QSOs. The other models are
variations on this set. Models L0.3, L2, L3, L3b and LX have the same
normalization of the power spectrum as L1, but models $L1\sigma .65$ and
$L1\sigma .4$ have smaller $\sigma_8$. For the standard CDM model S3 we
impose the normalization $\sigma_8=0.5$ deduced from the abundance of
galaxy clusters (Eke, Cole \& Frenk 1996). We have used \cmbfast
(Seljak \& Zaldarriaga 1996) to compute the appropriate linear matter
transfer functions.

These simulations follow the evolution of a periodic, cubic region of
the universe and are performed with a modified version of \hydra
(Couchman \etal 1995). \hydra combines hierarchical P3M gravity
(Couchman 1991) with smoothed particle hydrodynamics (SPH, Lucy 1977,
Gingold \& Monaghan 1977). One of us (TT) has parallelized this code
for the Origin 2000 at DAMTP, Cambridge. These simulations use $64^3$
particles of each species in a box of co-moving size 2.5$h^{-1}$Mpc, so
the SPH particle masses are $1.14\times 10^6\, (h/0.65)^{-3} M_\odot$
and the CDM particles are more massive by a factor $\Omega_{\rm
CDM}/\Omega_b$. The gravitational softening lenght is 5 kpc. All our
simulations were run with the same initial phases to minimize the
effects of cosmic variance when comparing the different models. We have
also run an additional model (L3b) which simulates a larger region
(5.0$h^{-1}$Mpc) with $128^3$ particles of each type and thus the same
numerical resolution as the other models. A simulation typically
requires 1000-1500 steps to reach $z=3$.

The detailed expressions for the heating and cooling rates as a
function of temperature and ionizing flux are taken from Cen (1992)
with some minor modifications (Theuns \etal 1998). We take the
photoheating rate in the optically thin limit from Haardt \& Madau
(1996). This heating rate, denoted as \lq HM\rq~ in the $\epsilon_{\rm
He}$ column of the Table, is imposed on models L1, L1$\sigma .65$ and
L1$\sigma .4$. Models L2 and L3 are identical to L1, except that we
have multiplied the helium photoheating rate by factors two and three
respectively (keeping the ionization rate constant). Model L3b has the
same heating rate as model L3. In model L0.3 we have reduced the helium
photoheating by a factor 3. Model LX is identical to model L1, except
that we have included Compton heating by the X-ray background as a
function of redshift $z$ at a rate (Madau \& Efstathiou 1999)
\begin{equation}
{\cal H}_X(z) = 1.251\times 10^{-31}\,(1+z)^{4.3}\,\exp(-(z/5)^2)\,n_e\,,
\end{equation}
where ${\cal H}_X$ is the heating rate in erg cm$^3$ s$^{-1}$ and $n_e$
is number density of free electrons in cm$^{-3}$. Finally, model S3 is
a standard CDM model, with the appropriate Haardt \& Madau (1996)
helium heating rate increased by a factor 3.

For each model we compute spectra along 1200 random lines of sight
through the simulation box at a given redshift (usually $z=3$) and then
scale the ionizing background flux by the factor $A$ in the analysis
stage such as to give a chosen effective optical depth $\tau_{\rm
eff}$. The factor $A$ required to give $\tau_{\rm eff}=0.33$ at $z=3$
is given in Table~\ref{table:runs}. We process these simulated spectra
using the following procedure, designed to introduce the same biases as
present in observational data from the HIRES spectrograph on the Keck
telescope. Each spectrum is first convolved with a Gaussian with full
width at half maximum of FWHM = 8 \kms and re-sampled onto pixels of
width 3 \kms. Photon and pixel noise is added such that the total
signal-to-noise is 50.  The \lq continuum\rq~ of the spectra is then
fitted with the method described in Theuns \etal (1998).

We also compare flux statistics for our simulated spectra to an
observed HIRES spectrum of QSO 1422+231, kindly provided to us by
W. Sargent and M. Rauch. The \lya forest of this QSO extends over a
significant redshift range, over which the effective optical depth
evolves significantly. In addition, the signal-to-noise ratio varies as
a function of wavelength and flux. We model these effects for this
particular spectrum using the following detailed procedure (see also
Rauch \etal 1997). We begin by dividing the spectrum in two halves.
For each half, we take the simulation output at the appropriate
redshift and scale the background flux such as to match the observed
mean effective optical depth. We convolve the spectrum with a Gaussian
with full width at half maximum of FWHM = 6.6 \kms and re-sample it to
pixels of the same size as the observed spectrum. We calculate the
noise properties of the QSO spectrum as a function of flux 
and add Gaussian noise to the simulated spectra appropriate for
the flux in each pixel. In the observed spectrum we exclude regions
with identified metal absorption lines and the region close to the
emission redshift which might be influenced by the proximity
effect. Below we will show that this more ellaborate procedure produces
flux statistics in better agreement with the data than the simpler one
described earlier. We use the same automated version of \vpfit~
(Carswell \etal 1987) to fit Voigt profiles to both simulated and
observed spectra (see Theuns \etal 1998 for more details).

\begin{figure}
\setlength{\unitlength}{1cm}
\centering
\begin{picture}(7,9)
\put(-1, -3){\includegraphics{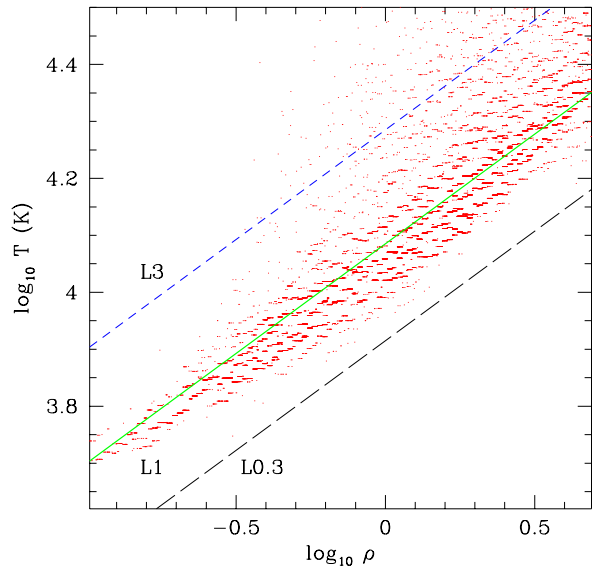}}
\end{picture}
\caption{Temperature-density relation for model L1 (points) and a
power-law fit (least absolute deviation; full line). Corresponding fits
for models L3 (short dashed) and model L0.3 (long dashed) are shown for
comparison. The \lq quantized\rq~ appearance of the data points close
to the full line is due to the usage of an interpolation scheme to
compute temperatures.}
\label{fig:eos}
\end{figure}

\begin{figure}
\setlength{\unitlength}{1cm}
\centering
\begin{picture}(7,9)
\put(-1., -3){\includegraphics{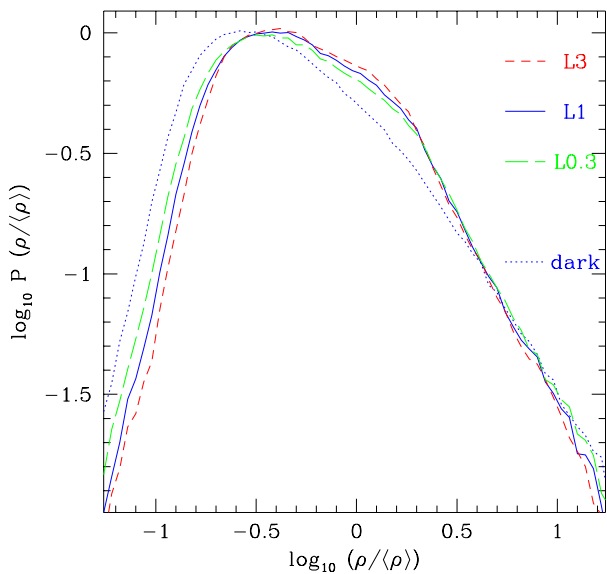}}
\end{picture}
\caption{Volume weighted gas density probability distribution for
models L0.3 (long dashed line), L1 (full line) and L3 (short dashed
line). Dark matter distributions are indistinguishable for these models
and are shown as the dotted line. Voids in the hotter model are less
empty because of increased Jeans smoothing, and so a smaller fraction
of the volume is occupied by very low density gas.}
\label{fig:prho}
\end{figure}

\begin{figure}
\setlength{\unitlength}{1cm}
\centering
\begin{picture}(7,9)
\put(-1., -3){\includegraphics{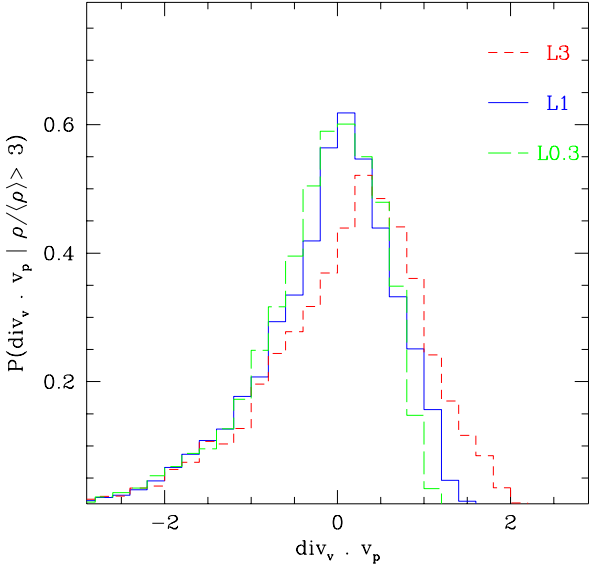}}
\end{picture}
\caption{Probability distribution for the divergence of the peculiar
velocity for models L0.3 (long dashed line), L1 (full line) and L3
(short dashed line). The larger pressure gradients in hotter models
lead to more pronounced expansion of the gas out of potential wells.}
\label{fig:pdiv}
\end{figure}

\section{Broadening of QSO absorbers}

In this section we illustrate the effect of different broadening
mechanisms on the widths of density peaks along a given line-of-sight.
In particular, we discuss the relative importance of the three
different thermal effects, namely Jeans smoothing, pressure-induced
peculiar velocities and thermal broadening and compare these to the
effect of the amplitude of dark matter fluctuations. In the following
sections we will investigate how the different broadening mechanisms
affect a number of different flux statistics.

\subsection{Thermal smoothing}

\subsubsection{The spatial distribution and the velocity of the gas} 

\begin{figure}
\setlength{\unitlength}{1cm}
\centering
\begin{picture}(7,12)
\put(-1.7, -3){\includegraphics{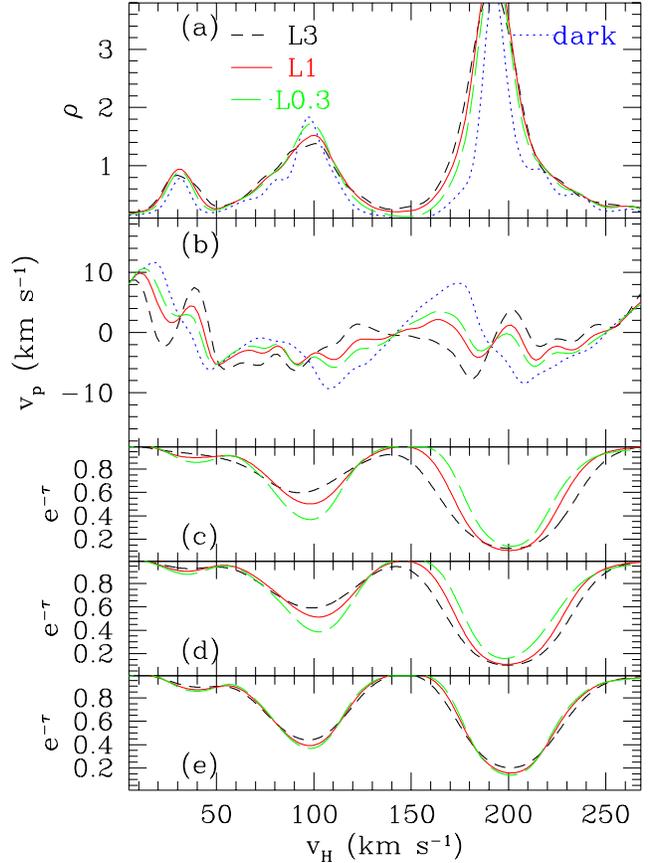}}
\end{picture}
\caption{Top panels show the density and peculiar velocity along a
sight line picked from the simulation, for models L0.3 (long dashed
line), L1 (full line) and L3 (short dashed line). The dark matter
profiles for these models are identical and are shown dotted. Panel (c)
shows the corresponding spectra. Panel (d) shows spectra for the case
where peculiar velocities are not included. Finally, panel (e) shows
the spectrum for model L0.3, but computed with the temperature-density
relation of models L0.3 (long dashed line), L1 (full line) and L3
(short dashed line). In the hotter models, density peaks are broader,
expansion velocities higher and the corresponding absorption features
wider.}
\label{fig:spec}
\end{figure}

Figure~\ref{fig:eos} shows the temperature-density relations for the
LCDM models where the HeII photo-heating is (0.3,1,3) times that in the
optically thin limit, respectively.  Over the interval shown, this
relation is well approximated by a power law. The slope of this power
law is determined by the reionization history (Hui \& Gnedin 1997),
which is the same for all models plotted here. The temperature $T_0$ at
the mean density for $z=3$ is also given in Table~\ref{table:runs}.  As
expected, $T_0$ increases with the heating rate as $T_0\propto
\epsilon_{\rm heat}^{1/1.7}$ (Miralda-Escud\'e \& Rees 1994).

In our models, baryons represent less than 15 per cent of the mass and
increasing their temperature does not affect the dark matter
distribution very much. In contrast, the gas distribution itself
changes significantly as the temperature is raised, due to the larger
pressure gradients that arise between high and low density regions.
The increased pressure gradient forces the gas to settle into a
configuration with a smaller density gradient in order to reach
pressure equilibrium ($\nabla P /\rho = - \nabla \phi$), leading to
broader structures. This effect is illustrated in Figure~\ref{fig:prho}
where we plot the probability distribution function (PDF) of the gas
density for the three models. Hotter models have a smoother density
distribution with less empty voids and less gas at high density (and
hence a narrower PDF), while the PDF of the DM density is
unchanged. (Note that the shift between the PDFs of gas and DM shows
explicitly that our simulations resolve the Jeans length.)

Another consequence of larger pressure gradients is that
pressure-induced peculiar velocities in the gas become larger (see also
Bryan et al. 1999). Figure~\ref{fig:pdiv} shows the PDF of the
divergence of the velocity field for the gas at an overdensity $\ge
3$. The distributions shift towards more positive values of ${\rm div}\
v_{\rm p}$, \ie a larger fraction of the gas at the center of sheets is
expanding, for hotter models. Figure~\ref{fig:spec}, to be discussed in
more detail below, shows explicitly for a typical sight line how
pressure-induced forces generate peculiar velocities that lead to a
smoothing of the gas with respect to the underlying dark matter.

The temperature also affects the normalization constant which
determines the mean flux level for a given ionizing background. If the
gas distribution were to be independent of temperature, then this
constant would simply scale as $T^{-0.7}$ due to the temperature
dependence of the recombination coefficient. There is, however, less
gas locked in dense regions in the hotter models and consequently the
temperature dependence is slightly weaker.

\subsubsection{The widths of QSO absorbers}

Figure~\ref{fig:spec} shows the density and velocity fields and the
corresponding \lya spectrum along a typical sight line for different
models. Panel (a) shows how a higher temperature leads to broadening of
the density peaks in real space. The density peaks in the gas
distribution are also significantly broader than in the underlying dark
matter due to this Jeans smoothing. Panel (b) shows the corresponding
peculiar velocity fields. The gas exhibits significant velocity bias:
whereas the dark matter is collapsing towards density peaks, the gas is
infalling well away from the central peak but is {\em expanding} near
the density maxima (see also Bryan et al. 1999).  Photoheating is in
the process of redistributing gas in the shallow potential wells where
it collected at higher redshift when the temperature was lower. This
effect is obviously stronger for models with a higher heating rate.

The combination of thermal broadening, Jeans smoothing and pressure
induced peculiar velocities makes absorption features broader for
hotter models, as shown in panel (c). The relative contribution of each
of these effects is shown in the final two panels. Panel (d) shows the
case where no peculiar velocities are included in the computation of
the simulated spectra. Comparing panels (c) and (d) we see that both
peculiar velocities and Jeans smoothing are important in determining
the shape of the line wings.  The wings of lines in model L3 are
broader than those for model L0.3 in panel (d) ({\em without} peculiar
velocities). The difference is even greater in panel (c) (\ie with
peculiar velocities). Finally, panel (e) shows spectra for a given
density distribution and peculiar velocity field (model L0.3), but with
different temperature-density relations imposed. These spectra which
now only differ in the amount of thermal broadening look much more
similar, but still the shape of the absorption profiles near the peak
are more rounded in the hotter models. Note that there are also some
differences in the line wings.

\subsection {The effect of small scale power}
\label{sect:sspower}

Hui \& Rutledge (1999) derived a theoretical expression for the shape
of the $b$-distribution that fits the observations fairly well. They
made the {\em Ansatz} that QSO absorption lines arise from peaks in the
density field. Their model explains the observed power law tail of
large $b$-values as being due to the frequent low-curvature
fluctuations whereas the sharp cut-off at low $b$ is a consequence of
high curvature peaks being exponentially rare. The model also predicts
that the widths of the absorption features should depend on the
amplitude of the underlying dark matter fluctuations. Models with
little small scale power are expected to have on average broader
density peaks and hence broader absorption lines.

\begin{figure}
\setlength{\unitlength}{1cm}
\centering
\begin{picture}(7,12)
\put(-1.7, -3){\includegraphics{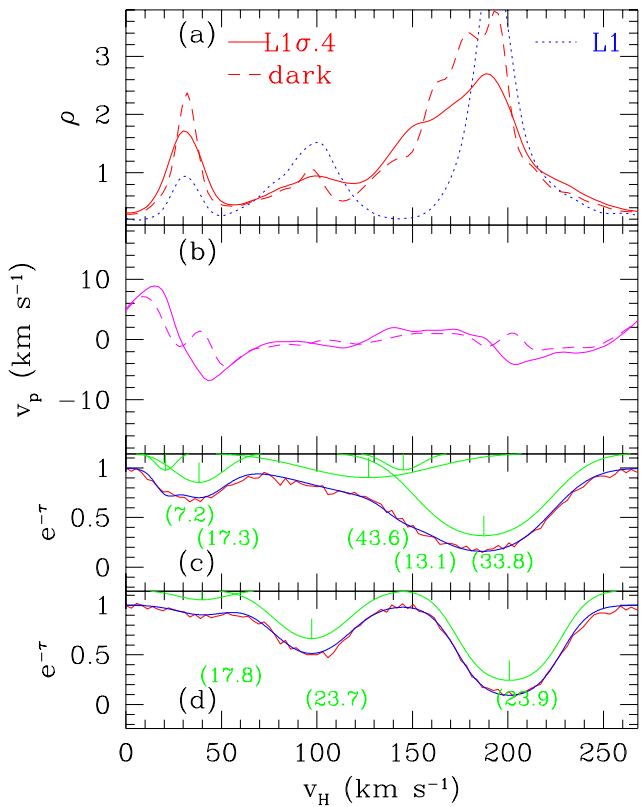}}
\end{picture}
\caption{Same sight line as shown in figure~\ref{fig:spec} for the low
fluctuation amplitude model L1$\sigma$.4. Top two panels show density
and peculiar velocity, for the dark matter (dashed line) and gas (full
line). The gas density for model L1 is shown dotted in panel (a) for
comparison. Panel (c) shows the corresponding fitted spectrum (with
noise added, wiggly line), the fit obtained using \vpfit~ (full line),
and the individual Voigt components (full lines with vertical line at
maximum absorption), off-set vertically for clarity. The numbers in
brackets denote the width of the fitted Voigt profile in km s$^{-1}$
for the lines with $\NHI > 10^{12.4}$ cm$^{-2}$. Panel (d) shows the same
as panel (c) for model L1.}
\label{fig:spec0.4}
\end{figure}

Figures~\ref{fig:spec} and ~\ref{fig:spec0.4} demonstrate that the
density peaks are indeed wider in the extreme low $\sigma_8$ model
L1$\sigma .4$ ($\sigma_8=0.4$) than in model L1 ($\sigma_8=0.9$, both
models have almost identical temperatures). The collapse of the density
fluctuations around 100 and 200 km s$^{-1}$ is significantly further
advanced in model L1 compared to model L1$\sigma .4$ and the collapsed
structures are much less extended in velocity space. As a result the
absorption features in model L1$\sigma .4$ are much broader than in L1.

However, as is also shown in the figure, the number of Voigt components
fitted to the resulting spectra by the Voigt profile fitting routine
\vpfit~ are different: the spectrum of L1$\sigma$.4 is fitted with five
components while that of L1 is fitted with only three components. It is
therefore not obvious that standard Voigt profile fitting routines
should require larger $b$-parameters to fit the more extended density
peaks in models with smaller amplitude of dark matter fluctuations
where the peaks often show significant substructure. We will show in
detail in the next section that this is generally not the case.

\section{The Doppler parameter distribution}
\label{sect:voigt}
\begin{figure}
\setlength{\unitlength}{1cm}
\centering
\begin{picture}(7,14)
\put(-2.5, -5){\includegraphics{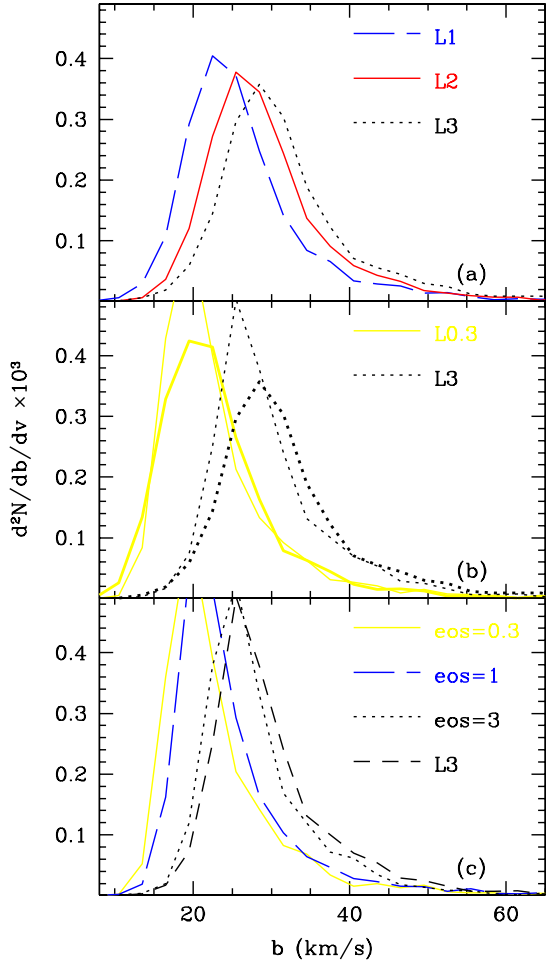}}
\end{picture}
\caption{Number of absorption lines per unit width $db$ and per
interval $dv$ in Hubble velocity as a function of line width. Included
are lines with column density in neutral hydrogen $N_\h>10^{13}$
cm$^{-2}$ for which \vpfit~ gives an estimated error in $b$, $b_{\em
err}<0.25 b$. Panel (a) compares models with increased helium heating
L1-L3. Panel (b) compares models L0.3 and L3 with (thick lines) and
without (thin lines) peculiar velocities. Finally, panel (c) compares
model L0.3 without peculiar velocities with the temperature-density
relations of model L0.3, L1 and L3 imposed. Model L3, without peculiar
velocities, is shown as the short dashed line.}
\label{fig:bdist}
\end{figure}

We compute the Doppler parameters from Voigt profiles fits to
absorption features. For both our simulated spectra and the observed
spectrum of the $z_{\rm em}=3.62$ QSO 1422+231, we use the same
automated version of \vpfit~.

\subsection{The global shift of the distribution with temperature}

The general shape of the $b$-parameter distribution is rather skewed,
with a slow fall of in the number of lines larger than the mode, but a
rapid drop in the fraction of lines narrower than the mode. We will
refer to the position at small $b$ where the fraction of lines starts
dropping rapidly as the \lq cut-off\rq~. (A more rigorous definition is
possible (Schaye \etal 1999a) but here we will only compare
$b$-distributions between different models and so we do not need a more
accurate definition of cut-off, since the differences amongst the
models are relatively large anyway.)

The effect of increased heating on the $b$-parameter distribution is
shown in figure~\ref{fig:bdist}a. Increasing the helium heating rate
(models L1 to L3) shifts the cut-off in narrow lines from $\le 20$\kms
to $\ge 25$\kms and decreases the fraction of lines at the peak. In
addition to that, the hotter model L3 has significantly more broad
lines with $b\ge 40$\kms than L1. Note that we have not normalized the
distributions to demonstrate that the increase in broad lines is not
due to a decrease in the number of narrow lines. As discussed in the
previous section, the global shift is due to the combined effect of
Jeans smoothing, pressure-induced peculiar velocities and thermal
broadening.

The effect of peculiar velocities on the Doppler parameter distribution
has been controversial and both a narrowing and broadening has been
claimed. Figure~\ref{fig:bdist}b compares the line-widths distribution
for models L0.3 and L3 with and without peculiar velocities. The
effects of peculiar velocities are more pronounced for the hotter model
L3. Peculiar velocities broaden the absorption lines and also
significantly decrease the number of lines around the maximum in the
distribution. This behaviour can be understood by looking again at
figure~\ref{fig:spec}, which shows that pressure-induced peculiar
velocities tend to oppose the infall into sheets, and this effect is
larger for the hotter model. Consequently, peculiar velocities do not
change the distribution of $b$-parameters very much in the cold model
while in the hot model the absorption lines become broader.

Note that these pressure induced peculiar velocities are not due to
shocking of the gas: the higher temperature are purely a result of
increased photo-heating. It is photo-heating that drives the higher
density gas into its lower density surroundings, which results in the
changes in the density and velocity divergence PDFs shown earlier. The
high density gas collected at the bottom of the dark matter potential
wells at higher redshifts, when the heating rates were lower and the
gas was colder.

To illustrate the effect of pure thermal broadening, we have computed
spectra from the numerical simulations after having artificially
changed the temperature-density relation of the gas. The resulting
Doppler parameters for model L0.3, without peculiar velocity
contributions, and with the temperature-density relations of model
L0.3, L1 and L3 imposed are shown in
Figure~\ref{fig:bdist}c. Increasing the amount of thermal broadening
shifts the cut-off at small $b$ to larger values. The position of the
cut-off is thus a measure of the temperature of the gas. This is at the
heart of the method developed by Schaye et al. (1999a), who used
numerical simulations to demonstrate that the IGM temperature can be
measured to high accuracy by comparing the lower cut-off of the Doppler
parameter distribution of simulated spectra to that of the observed
distribution. Comparing the full line with the dotted line shows that
with increased thermal broadening the number of broad lines increases
as well.

The effect of Jeans smoothing can be seen by comparing the dotted with
the short dashed lines. Both these models are with peculiar velocities
set to zero and with the $T-\rho$ relation of model L3, but the first
uses the density distribution of L0.3 and the second the one of L3. The
extra broadening of the lines is thus purely due to the widening of the
absorbers in real space, due to the increased pressure gradients in the
hotter model. The wider structures have an increased differential
Hubble flow resulting in wider absorption lines. Note that the
magnitude of this effect is smaller than that of thermal broadening.

\subsection{The effect of small scale power}

\begin{figure}
\setlength{\unitlength}{1cm}
\centering
\begin{picture}(7,5)
\put(-3, -5.4){\includegraphics{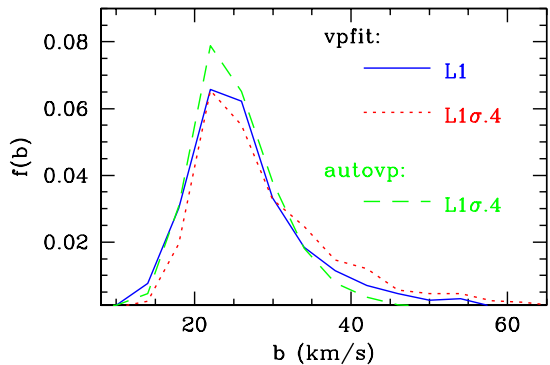}}
\end{picture}
\caption{Fraction of lines with a given width, $f(b)$, for models L1
(full line) and L1$\sigma .4$ fitted with \vpfit~ (dotted line) and
using \autovp~ (dashed line). Although the structures in the extreme low
$\sigma_8$ model are significantly broader than in L1, this is not
reflected in the $b$-distribution. The $b$-distributions obtained using
\vpfit~ and \autovp~ are very similar.}
\label{fig:fig3}
\end{figure}

In section~\ref{sect:sspower} we showed that models with less small
scale power have significantly wider absorption features. The effect of
this widening on the $b$-distribution (i.e., the probability
distribution of lines as a function of their width, here for lines with
$\NHI \ge 10^{13}$ cm$^{-2}$) is shown in figure~\ref{fig:fig3}, which
compares models L1 and L1$\sigma .4$, which have the same $T-\rho$
relation, but different values of $\sigma_8$. The two models have
almost identical $b$-parameter distributions, which can be understood
as follows. When fitting a Voigt profile to a relatively broad
absorption feature with a non-Voigt shape, as often occurs for models
where the broadening is not thermally dominated, there is a trade-off
between fitting a single broad Voigt line (which might not fit
particularly well) or fitting multiple narrower lines. \vpfit~ was
developed to \lq deblend\rq~ absorption features into thermally
broadened Voigt profiles and generally selects the option of fitting
more but narrower lines with thermal shapes. Therefore, even for the
wide absorption features in model L1$\sigma$.4 the widths of the fitted
Voigt profiles are still closely related to their ``thermal''
width. The $b$-distribution for models L1 and L1$\sigma$.4 are thus
very similar, since they have the same temperature.

We have fitted the two models with different $\sigma_8$ with another
Voigt profile fitting routine, \autovp, kindly provided to us by Romeel
Dav\'e (Dav\'e \etal 1997) to check if this result is special to
\vpfit~. The $b$-distribution obtained with \autovp~ is very similar to
that obtained with \vpfit~, in particular, there is no strong dependence
of the $b$-distribution on the amplitude of the dark matter
fluctuations. This result contrasts with the findings of Bryan \&
Machacek (1999), who did find a strong dependence of the
$b$-distribution on $\sigma_8$ using different profile fitting
software.

\begin{figure}
\setlength{\unitlength}{1cm}
\centering
\begin{picture}(7,5)
\put(-3, -5.4){\includegraphics{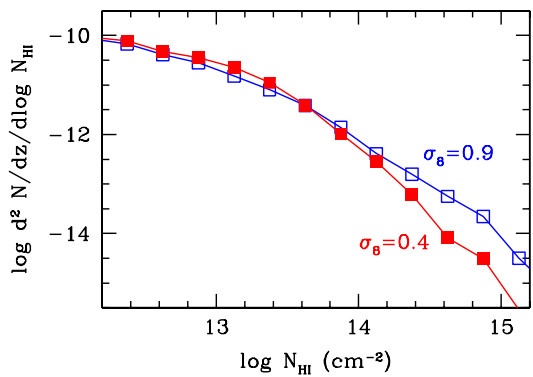}}
\end{picture}
\caption{Column density distribution for model L1 (open squares)
compared to L1$\sigma$.4 (filled squares). The low normalization model
has significantly less high column density systems.}
\label{fig:fig3b}
\end{figure}

Finally, we confirm the result of Gnedin (1998) that changing the
amplitude of dark matter fluctuations affects the column density
distribution as shown in figure~\ref{fig:fig3b}.  At low column
densities this is because many more weak lines are now required to fit
the broader features. At the same time, L1$\sigma$.4 has fewer strong
lines and so the column density distribution steepens with decreasing
$\sigma_8$.


\subsection{Comparison to the observed Doppler parameter distribution}
\begin{figure}
\setlength{\unitlength}{1cm}
\centering
\begin{picture}(7,10)
\put(-3, -5.4){\includegraphics{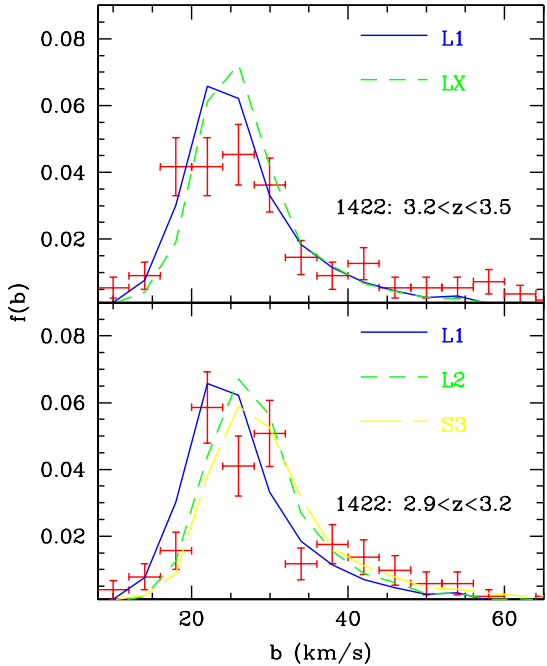}}
\end{picture}
\caption{Fraction of lines with a given width, $f(b)$, versus $b$ for
lines with $\NHI > 10^{13}$ cm$^{-2}$ with formal errors $b_{\rm
err}\le 0.25b$, obtained from \vpfit~. Symbols with (Poissonian) error
bars are from fitting the spectrum of QSO 1422+231 using our automated
version of \vpfit~, which produced $\sim 350$ lines that met our
selection criterion. Top (bottom) panel contains the top (bottom) half
of the observed spectrum, the corresponding redshift ranges are
indicated. The top panel compares the data to models L1 and LX, the
bottom panel to L1, L2 and S3. L1 reproduces the observed distribution
well for the top half of QSO 1422+231, but the bottom half seems to be
fit better by L2. Models L1 and LX are very similar, as are models L2
and S3.}
\label{fig:bdistdata}
\end{figure}

Figure~\ref{fig:bdistdata} compares the Doppler parameter distribution
of some of our models to the observed distribution for QSO 1422+231. We
have fitted the \lya-spectrum of this QSO using our automated version
of \vpfit, thereby analysing observed and simulated spectra in the same
way. We plot the $b$-distribution of the fitted lines for the top and
bottom half of the spectrum separately.

Model L1 reproduces the observed $b$-distribution of the top half of
QSO 1422+231 reasonably well, providing a good fit to both the cut-off
at small $b$, and to the tail of broader lines with $b\ge 40$ km
s$^{-1}$. However for the lower redshift half, L1 seems to cut-off at
too small values of $b$, and in addition fails to reproduce the
observed number of broader lines. Interestingly, the hotter model L2
{\em does} provide a good fit to the bottom half of QSO 1422+231 over
the whole $b$ range. A more thorough investigation (Schaye \etal 1999b)
demonstrates that the low redshift half of QSO 4122+231 is indeed
hotter.

Turning our attention to a comparison between the different models, we
see that models L1 and LX produce very similar
$b$-distributions. Compton heating by the hard X-ray background is not
able to raise the temperature of the gas probed by these lines
significantly above the value which results from helium photoheating as
computed by Haardt \& Madau (1996). However there is a significant
difference in temperature in regions of very low densities between L1
and LX, which might be measurable with other flux statistics. Finally,
models L2 and S3 also have very similar $b$-distributions. Both models
have a temperature of $\ga$ 15000 K at mean density.  In the standard
CDM model S3 ($\Omega_{\rm m}=1$), the adiabatic cooling time scale is
smaller than in the $\Lambda$CDM model ($\Omega_m=0.3,\,
\Omega_\Lambda=0.7$). This explains why the Doppler parameter
distribution are similar despite the larger HeII heating rate in model
S3.

\section{Other flux statistics}
\label{sect:other}

In this section we investigate the effect of the thermal broadening
mechanism and that of the amplitude of dark matter fluctuations on
small scales on the the one- and two-point functions of the flux and on
the flux power spectrum. These statistics are in principle easier to
compute than the $b$-distribution but are likely to be affected by a
number of observational biases. For example, Rauch et al. (1997) showed
that the shape of the one-point function (flux probability
distribution) evolves strongly with redshift, as a consequence of the
rapid evolution of the effective optical depth. Therefore, given that a
QSO spectrum spans a relative large redshift range, such evolution
should be accounted for in a comparison with simulations. We have
followed the procedure described in detail in Section~2 to compare our
simulations to the spectral statistics of QSO 1422+231.

\subsection{The one-point function of the flux} 
\label{sect:1pt}

\begin{figure*}
\setlength{\unitlength}{1cm} \centering
\begin{picture}(21,7)
\put(-2.5, -7){\includegraphics{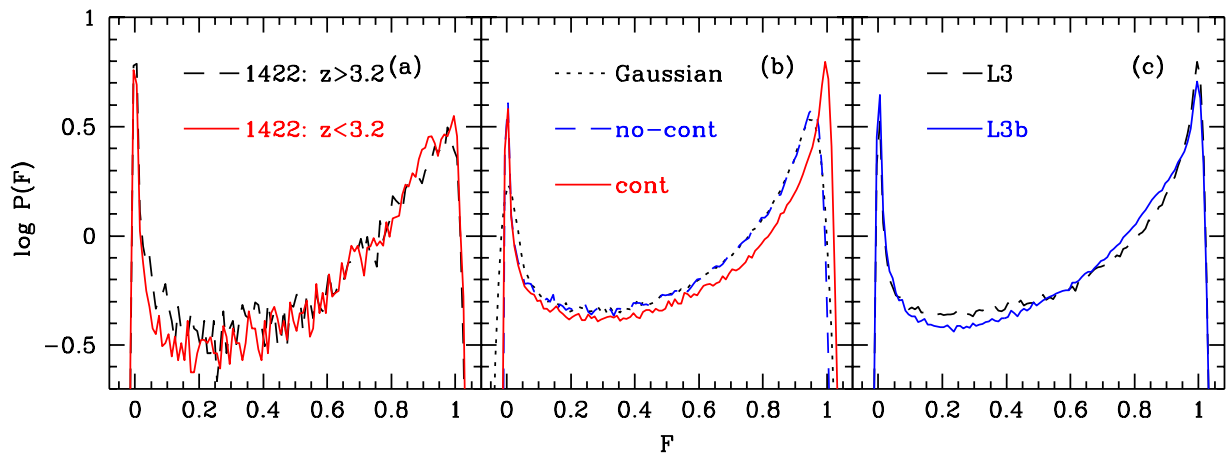}}
\end{picture}
\caption{One-point function of the flux. Panel (a): comparison of top
half of the \lya forest of QSO 1422+231 (dashed line) with the bottom
half (full line). Cosmic variance, optical depth evolution and varying
signal to noise are responsible for the differences. Panel (b):
simulated spectra at the corresponding redshift, assuming Gaussian
noise with effective signal-to-noise of 50 and no continuum fitting
(dotted line) and using the noise properties of QSO 1422+231 without
(short dashed line) and with (full line) continuum fitting. Continuum
fitting affects the high end significantly, whereas the noise
properties are important at low flux levels. Panel (c): comparison of
models L3 (dashed line) and L3b (full line) illustrating the effect of
missing large scale power.}
\label{fig:one-point1}
\end{figure*}

\begin{figure*}
\setlength{\unitlength}{1cm} \centering
\begin{picture}(21,7)
\put(-2.5, -7){\includegraphics{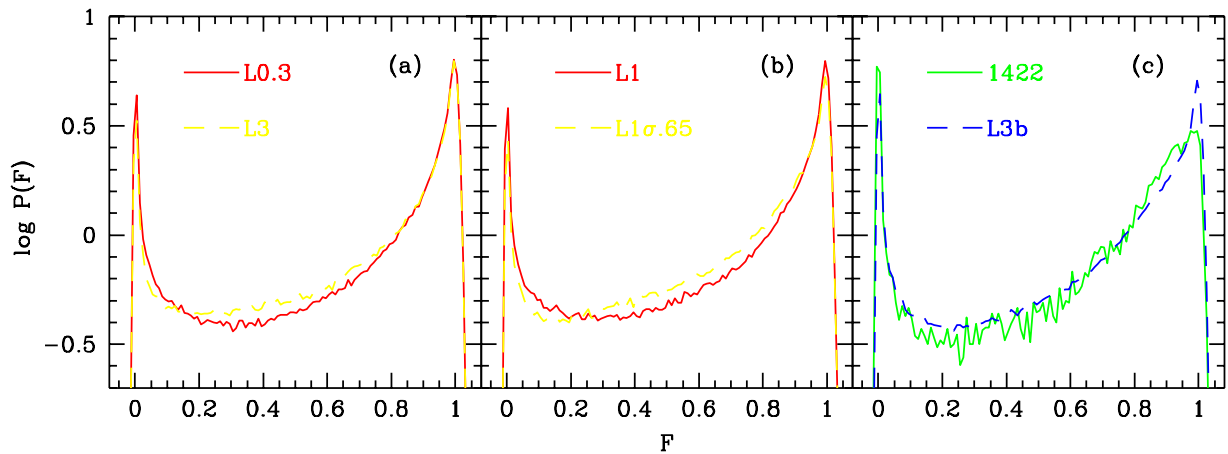}}
\end{picture}
\caption{Comparison between one-point functions. In panel (a) model
L0.3 (full line) versus L3 (dashed line); in panel (b): model L1 (full
line) versus model L1$\sigma .65$ (dashed line); in panel (c) model L3b
(dashed line) versus QSO 1422+231 (full line). Increasing temperature
has the same effect as decreasing $\sigma_8$. Model L3b reproduces the
observed one-point function reasonably well.}
\label{fig:one-point2}
\end{figure*}

The shape of the one-point function depends strongly on the assumed
mean absorption, the continuum fit, and on the noise properties.  For
example, figure~\ref{fig:one-point1}a shows the differences between the
top and bottom halves of the \lya spectrum of QSO 1422+231 due to
redshift evolution of the mean absorption, cosmic variance, and
non-uniform signal to noise over the spectrum.

The importance of using the correct noise properties is illustrated in
figure~\ref{fig:one-point1}b where we compare the one-point function
for a simulated spectrum with Gaussian noise independent of flux
(dotted line) and with the more ellaborate method described in
Section~2 (short dashed line). There are significant differences,
especially for saturated pixels ($F=0$).  Adding noise broadens the
peak around nearly saturated pixels considerably compared to the case
where the noise properties are similar to that in the observed
spectra. In addition, a relatively large fraction of the pixels around
$F=0$ comes from a few rare high column density systems. To test this
we have removed the four highest column density systems from the
spectrum of Q1422 (column densities between $10^{15.2}$ and $10^{17.9}$
cm$^{-2}$) and recomputed the one-point function. These two functions
now differ significantly around $F=0$. Since such strong systems are
relatively rare, the shape of the one-point function round $F=0$
suffers from small number statistics.

For the full curve in figure~\ref{fig:one-point1}b the spectra were
calculated as before but we have now also continuum fitted the
simulated spectra as described in Theuns et al. (1998). Comparing the
short dashed line with the full line shows that, not surprisingly,
continuum fitting has a big effect on the shape of the one-point
function around $F=1$. The smallness of our simulation box probably
exacerbates this problem but it should be kept in mind that the
continuum level of observed QSOs is often rather ill defined,
especially at higher $z$. In the remainder of this section, we will
present one-point functions computed using the noise properties of QSO
1422, and including continuum fitting. Finally,
figure~\ref{fig:one-point1}c compares models L3 and L3b: missing large
scale power has a relatively significant effect on the one-point
function, but this should not limit comparison between different models.

Figure~\ref{fig:one-point2}a compares the one-point function for the
cold model L0.3 to the hot model L3. The trend is as expected: the cold
model has relatively more pixels round $F\sim 0.1$, since a larger
fraction of the colder gas has collapsed to higher densities, but less
absorption around $0.2\le F\le 0.7$, since the hotter model is denser
in the regions surrounding dense peaks. This is due to the increased
pressure forces in the hotter model driving more of the gas out of the
centres of potential wells into the lower density regions. The middle
panel shows that the effect of less small scale power is similar to
that of higher temperatures: L1$\sigma$.65 has fewer pixels round $F=0$
but more absorption in the low density regions around $0.2\le F\le
0.7$. Because of the difference in power, gas in L1$\sigma$.65 is less
collapsed in high density regions, and so there is correspondingly more
gas in lower density regions to boost the one-point function in the
region $0.2\le F\le 0.7$. Machacek \etal (1999) also showed that the
shape of the optical depth distribution is a sensitive probe of the
amplitude of density fluctuations on small scales. However, as we have
just demonstrated, the effect of temperature on the flux decrement
distribution and that of the fluctuation amplitude of the DM density
are largely degenerate. Finally, figure~\ref{fig:one-point2}c shows
that model L3b reproduces the observed one-point function reasonably
well.

\subsection{The two-point function of the flux}

The two-point function of the flux, $P(F_1,F_2,\Delta v)$, is the
probability that two randomly chosen pixels separated by a velocity
interval $\Delta v$ will have transmitted fluxes $F_1$ and $F_2$
(Miralda-Escud\'e \etal 1997). A convenient way of visualizing this function
is by considering the mean flux difference for $F_1$ in a given flux
interval, that is, select a flux interval $\delta F_1$ and consider
the mean flux difference
\begin{eqnarray}
&\overline{\Delta F}&(v;\delta F_1)\nonumber\\
&\equiv&\!\!\!\!\!\!\!
\int_{\delta F_1} \left\{\int_{-\infty}^{+\infty} (F_1-F_2)
P(F_1,F_2,\Delta v) dF_2\right\} dF_1/\delta F_1\,.
\end{eqnarray}

The shape of $\overline{\Delta F}(v;\delta F_1)$ contains information
on the profile of the absorbers, or in the traditional language, on the
widths of the lines. For large velocity differences, the correlation
between the pixels is lost and the two-point function contains no extra
information over the one-point function. In this limit, the mean flux
difference will tend to the mean flux minus the mean flux of the pixels
in the flux interval under consideration. Since $\overline{\Delta F}$
is an averaged quantity, the interpretation of the two-point function
is not always obvious.

\begin{figure*}
\setlength{\unitlength}{1cm}
\centering
\begin{picture}(21,7)
\put(-2.5, -7){\includegraphics{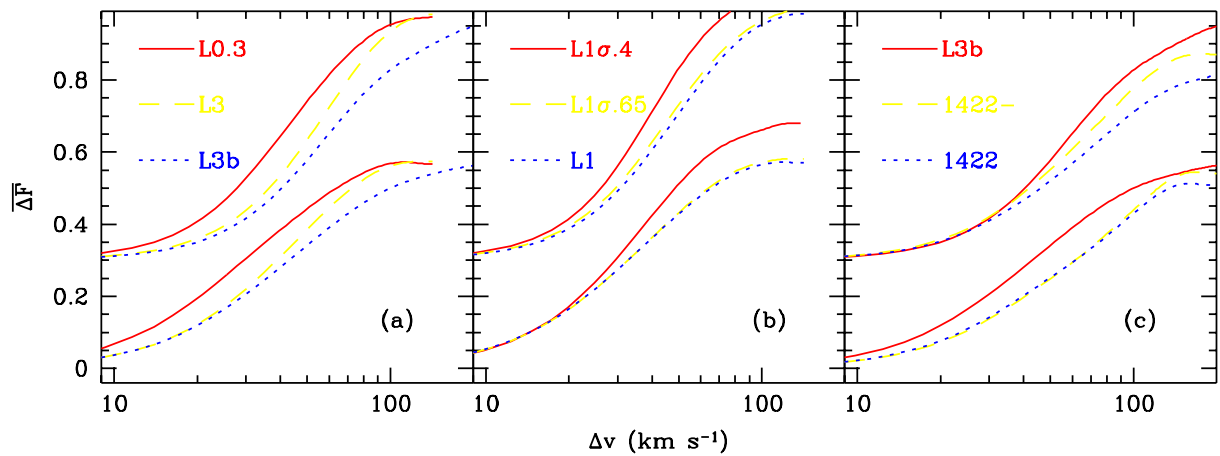}}
\end{picture}
\caption{Two-point functions for several models and for QSO 1422+231
for flux class 1 ($-0.05 < F_1< 0.05$) and flux class 2 ($0.05 < F_1
< 0.15$). Panel (a): models L0.3 (full line), L3 (dashed line) and L3b
(dotted line). Top curves are for class 1 and are off-set vertically by
0.3. Panel (b) is the same as (a) but for models L1$\sigma$.4 (full
line), L1$\sigma$.65 (dashed line) and L1 (dotted line). Panel (c) is
the same as panel (a) but for model L3b (full line), QSO 1422 with the
four strongest lines removed (dashed line) and QSO 1422 (dotted
line). Panel (a) shows that temperature and box size have an
important effect on the two-point function, in contrast to the amount
of small scale power (panel b). There are significant differences
between the simulated and observed two-point functions, but these are
at least partly due to missing large scale power in the simulations
(panel a), and to strong absorbers in the observations (panel c).}
\label{fig:two-point}
\end{figure*}

Figure~\ref{fig:two-point} compares the two-point function for several
models and for QSO 1422 and for two flux intervals, $-0.05\le F_1\le
0.05$ (flux class 1) and $0.05\le F_1\le 0.15$ (class 2). The
temperature of the gas is clearly important: the two-point function
shows that pixels are correlated over larger scales for hotter models,
as expected given the broadening of absorption lines with increasing
temperature. As with the one-point function, there is also a rather
important effect of missing long waves in our small simulation
boxes. The effect of small scale power on the other hand is
surprisingly small, the two-point functions of models L1 and
L1$\sigma$.65 are almost indistinguishable.  Only for model
L1$\sigma$.4 which has very much less small scale power are the
differences significant. This is mostly due the much lower number of
high column density absorbers in this low normalization model.

Model L3b reproduces the two-point function of QSO 1422 up to 30\kms,
but then appears less correlated for larger velocity differences than
the data. Unfortunately, just as for the one-point function, flux class
1 is relatively strongly influenced by the presence of a few strong
absorbers in the data (compare the dashed with the dotted 
line). The relatively good correspondence between model L3b and the
data suggests that the temperature of the IGM at mean density is high,
$\ga 15000$K at this redshift.

\subsection{Flux power spectrum} 

\begin{figure*}
\setlength{\unitlength}{1cm}
\centering
\begin{picture}(21,7)
\put(-2.5, -7){\includegraphics{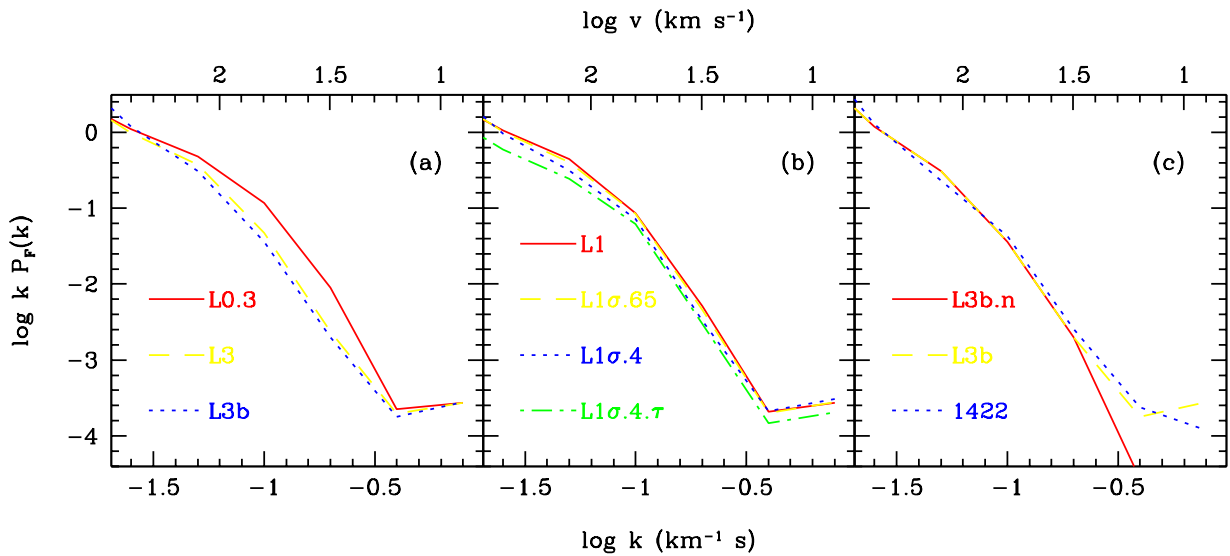}}
\end{picture}
\caption{Flux power spectrum for several models and for QSO
1422+231. Panel (a): models L0.3 (full line), L3 (dashed line) and L3b
(dotted line); panel (b): models L1 (full line), L1$\sigma$.65 (dashed
line) and L1$\sigma$.4 (dotted line). Model L1$\sigma$.4 but with the
same ionizing background as L1 (same factor $A$, see
table~\ref{table:runs}) is shown with the dot-dashed line. Panel (c)
shows models L3b without (full line) and with noise (dashed line), and
QSO 1422 (dotted line).  The bottom axes refer to wave number $k$, the
top axes to the corresponding Hubble velocity $v=2\pi/k$. Panel (a)
shows that temperature has an important influence on the shape of
$P_F(k)$, and that the effect of missing long waves is small. Panel (b)
shows that $\sigma_8$ has relatively little effect, if models are
compared at the same mean flux. Panel (c) shows that the shape of
$P_F(k)$ for $k\ge -0.7$ is set by the noise, and that L3b fits the
power spectrum of QSO 1422 on small scales very well.}
\label{fig:powspec}
\end{figure*}

We begin by discussing the relation between the flux power spectrum and
the underlying spectrum of density fluctuations. For a Gaussian random
field, the power spectrum of density fluctuations along a sight line,
$P_{\rm 1D}(k)$, is defined as the ensemble average of the square of
the Fourier transform of the density $\rho(v)$,
\begin{equation}
\delta_\rho(k) = \int_0^V \rho(v) exp(-ikv) dv/\langle\rho\rangle\,,
\end{equation}
where $V$ is the length of the sight line, and $\rho$ is the density as
function of the Hubble velocity $v$. It is related to the three
dimensional power spectrum of the field, $P_{\rm 3D}(|\bfk|)$, through
an integration (Kaiser \& Peacock 1991)
\begin{equation}
P_{\rm 1D}(k) = {1\over 2\pi} \int_k^\infty P_{\rm 3D}(q) q dq\,.
\label{eq:pk}
\end{equation}
This equation shows that $P_{\rm 1D}(k)$ has contributions from all
waves with $q>k$. Since Jeans smoothing introduces a small-scale
cut-off in the power spectrum of the three dimensional density field,
it influences $P_{\rm 1D}(k)$ for the gas on all scales.

The observed flux is related to the gas density in redshift space by
the relation
\begin{equation}
F=  \exp[{-{\cal A} (\rho_{\rm gas}/<\rho_{\rm gas}>)^{\alpha}}]\,, 
\label{eq:fvr}
\end{equation}
and the flux power spectrum $P_F(k)$ is defined as the ensemble average
of the square of the Fourier transform of $F(v)$,
\begin{equation}
\delta_F (k) = \int_0^V F(v) \exp(-ikv) dv\,.
\end{equation}
Peculiar velocities and thermal broadening contribute to the small
scale cut-off of $P_F(k)$. The factor ${\cal A}$ ($\propto A$) in
equation~\ref{eq:fvr} is a normalization constant that determines the
mean flux, and acts as a \lq bias\rq~ factor between density and
flux. The exponent $\alpha$ depends on the slope of the $\rho$-$T$
relation. $P_F(k)$ differs from $P_{1D}(k)$ due to the combined effects
of Jeans smoothing, redshift space distortions and thermal broadening,
and in addition due to the non-linear relation~\ref{eq:fvr}.

Figure~\ref{fig:powspec} compares $k\,P_F(k)$ for several models and
for QSO 1422+231. On small scales still above the resolution limit in
the dark matter, the power spectrum dependence $P_{\rm 3D}\sim k^{-3}$
implies that $k\,P_{\rm 1D}(k)$ depends only weakly on $k$,
highlighting that the observed sharp cut-off in $k\,P_F(k)$ is due to
thermal smoothing. To compute these power spectra, we performed the
following steps. The simulated spectra were calculated in the same way
as we before, \ie, scaling the models to the same mean flux, and adding
noise with the same properties as that of the spectrum of QSO 1422. We
have chosen a somewhat smaller pixel size of $2.2$\kms to facilitate
the use of a fast Fourier transform routine to calculate $P_F(k)$. To
calculate the power spectrum of the observed spectrum, we re-sampled
the spectrum to the same pixel size as the simulated spectra and
divided it into stretches with lengths equal to that of our simulated
spectra. We took care to use streches that have a flux close to one at
the edges and in addition multiplied the flux with a windowing
function, in order to be able to use \lq periodic\rq~ boundary
conditions in the FFT. We then computed the Fourier transform for each
stretch separately and averaged the power per mode. To estimate the
influence of missing large scale power in our small simulation boxes,
we compare models L3 and L3b, which are identical expect for the size
of the simulated region. As we show below, the differences in the flux
power spectrum between these models is very small.

On large scales where gas follows dark matter, equations~\ref{eq:pk}
and \ref{eq:fvr} can be combined to infer the three dimensional power
spectrum from the flux power spectrum (Croft \etal 1997). In contrast,
on small scales, log$k$(\smk)$\ge$ -1.4, thermal smoothing strongly
suppresses the amplitude of $P_F(k)$. Panel (a) demonstrates how this
cut-off moves to smaller wavenumbers for higher temperatures. The solid
curve is for the cold model L0.3 while the dashed curve is for the hot
model L3. The dotted curve is for the larger box L3b and there is a
hint of increased power on the largest scales but the effect seems to
be small.

Panel (b) shows that there is little dependence of the small scale flux
power spectrum on $\sigma_8$. This may seem surprising at first but is
largely due to the fact that we have rescaled all simulated spectra to
the observed mean flux (see Croft \etal 1997 for a discussion of this
point). The dot-dashed line is for model L1$\sigma$.4, but imposing the
same normalization $A$ as for model L1, thereby increasing the
effective optical depth from $\sim .33$ to $\sim .43$. The power
spectrum for this model now differs significantly from the
others. Consequently, scaling models to the same mean flux removes most
of the dependence of the flux power spectrum on the fluctuation
amplitude of the DM distribution on these small scales. Given that the
shape of $P_F(k)$ depends strongly on temperature suggests it can be used
as an independent method to measure the temperature of the IGM.

Panel (c) compares L3b to the observed spectrum of QSO 1422. The full
line shows $k\,P_F(k)$ for L3b but without adding noise to the
spectra. Comparison to the dashed line shows that the shape of $P_F(k)$
for log$k$ (\smk)$\ge$-0.5 is set by the noise properties. Imposing the
noise of QSO 1422 onto our hot model L3b gives a simulated power
spectrum in excellent agreement with that of 1422. Therefore, both the
two-point function and the flux power spectrum on small scales suggest
that the IGM temperature at $z\sim 3.25$ is high, $\ga 15000$K.

\section{Discussion and conclusions}

We have used numerical simulations to investigate what determines the
widths of \lya absorbers in QSO spectra. These absorbers arise when a
line of sight intersects a peak in the neutral hydrogen density, which
itself is due to gas falling into the sheet-like and filamentary
potential wells of the underlying dark matter distribution. The widths
of the absorbers are set by two very different processes. Firstly, the
amount of small scale power determines the widths of the dark matter
potential wells. Secondly, Jeans smoothing of the gas due to its high
temperature ensures it is distributed much smoother than the dark
matter on small scales. Finally, the corresponding absorption features
are wider than the absorbers in real space due to peculiar velocities and
thermal broadening.

The strong dependence on the amount of small scale power is due to the
fact that the linearly extrapolated fluctuation amplitude on the Jeans
scale is of order unity for the models of interest. For this reason,
the collapse of structures on scales comparable to the Jeans length is
less advanced in models with a lower amplitude of DM fluctuations.

The effect of temperature on the widths of peaks in the gas
distribution can be understood as follows. Larger temperature gradients
between high and low density regions, as occur in hotter models, lead
to increased pressure gradients that push gas out into the wings of the
underlying dark matter potential wells. Consequently, less gas
collapses into high density regions and voids in the density
distribution become less underdense, making the peaks wider. This
smoothing process is particularly efficient for photoionized gas
because it obeys a temperature-density relation where the gas
temperature increases with density, and this increases the pressure
gradient over its isothermal value.

We have further investigated what effect the different broadening
mechanisms have on the resulting absorption spectra.  We used the
Doppler parameter distribution as obtained with the standard
Voigt-profile fitting routines \vpfit~ and \autovp, the one- and
two-point functions of the flux, and the flux power spectrum to
characterize the absorption spectra. 

The Doppler parameter distribution is mainly determined by the combined
effect of the three thermal smoothing mechanisms, thermal broadening,
thermally driven peculiar velocities and Jeans smoothing. It shows
little dependence on the amplitude of dark matter fluctuations. \vpfit~
and \autovp~ are therefore rather succesful in deblending absorption
features into thermally broadened Voigt profiles. Individual Voigt
components often trace the substructure in broader density peaks. This
is the reason why the Doppler parameter distribution shows little
sensitivity to the overall width of the density peaks, which themselves
depend strongly on the amplitude of dark matter fluctuations.

The flux two-point function and the flux power spectrum on small scales
are also mainly determined by thermal broadening and show likewise no
strong dependence on the amplitude of dark matter fluctuations. This is
at least partially due to the fact that we have chosen to scale our
simulated spectra to match the mean observed flux.  The one-point
function of the flux, however, depends on temperature and normalization
of the DM fluctuation amplitude in a similar way. This degeneracy
arises because the PDF of the density depends mainly on the {\it rms}
fluctuation amplitude of the gas, which itself depends on both the
fluctuation amplitude of the DM density and on the scale on which the
gas distribution is smoothed relative to the DM distribution.

We conclude that the Doppler parameter distribution, the flux two-point
correlation function and the flux power spectrum on small scales are
sensitive probes of the temperature of the photoionized IGM, while the
one-point function of the flux can be used to probe the fluctuation
amplitude on small scales once the temperature is known.

When comparing our simulated spectra with a high resolution spectrum of
QSO 1422+231, obtained with the HIRES spectrograph on the Keck
telescope, we find that the Doppler parameter distribution, the
two-point correlation function and the flux power spectrum are best
matched by models that have a temperature of the IGM of $\ga$ 15000K at
mean density at $z=3.25$. This temperature is $\sim$ 50 per cent higher
than in our reference model L1 that used the heating rates calculated
for the Haardt \& Madau spectrum. (Note however, that in this paper we
have only varied the amplitude of the $\rho-T$ relation, and not its
slope. This prevents us from making stronger statements about the value
of $T_0$.) The flux one-point function depends both on temperature and
the level of small scale fluctuations. A $\Lambda$CDM model with this
temperature matches the observed one-point function reasonably well.
Once an accurate measurement of $T_0$ is available, it should be
possible to determine $\sigma_8$ accurately from the \lya forest.

\section*{Acknowledgments}
We thank M. Rauch and W. Sargent for providing us with the HIRES
spectrum of the QSO 1422+231. JS thanks the Isaac Newton Trust,
St. John's College and PPARC for support. We thank R. Carswell for
helping us with \vpfit. This work has been supported by the \lq
Formation and Evolution of Galaxies\rq~ network set up by the European
Commission under contract ERB FMRX-CT96086 of its TMR programme.
Research conducted in cooperation with Silicon Graphics/Cray Research
utilising the Origin 2000 supercomputer at DAMTP, Cambridge.

{}
\end{document}